\documentclass[aps,pre,twocolumn,superscriptaddress,superscriptreference]{revtex4-1}

\usepackage{amsmath,bbold,bm,amssymb,scalerel,mathtools}
\usepackage{graphicx}
\usepackage{color}
\usepackage{enumitem}
\usepackage{algorithm,algpseudocode}
\usepackage{multirow}
\usepackage{colortbl,booktabs}
\usepackage{placeins}
\usepackage[usenames,dvipsnames]{xcolor}

\usepackage[colorlinks, linkcolor=BrickRed, urlcolor=blue!50!black, citecolor=blue!50!black]{hyperref}

\newcommand\equalhat{%
	\let\savearraystretch\arraystretch
	\renewcommand\arraystretch{0.3}
	\begin{array}{c}
		\stretchto{
			\scalerel*[\widthof{=}]{\wedge}
			{\rule{1ex}{3ex}}%
		}{0.5ex}\\ 
		=%
	\end{array}
	\let\arraystretch\savearraystretch
}

\newcommand{\gj}[1]{\textcolor{black}{#1}}

\usepackage[normalem]{ulem}

\begin{document}

\title{Mobility, response and transport in non-equilibrium coarse-grained models}

\author{Gerhard Jung}
\email{gerhard.jung.physics@gmail.com}
\affiliation{Laboratoire Charles Coulomb (L2C), Université de Montpellier, CNRS, 34095 Montpellier, France}

\begin{abstract}

We investigate two different types of non-Markovian coarse-grained models extracted from a linear, non-equilibrium microscopic system, featuring a tagged particle coupled to underdamped oscillators. The first model is obtained by analytically ``integrating out'' the oscillators and the second is \gj{based on a derivation} using projection operator techniques. We observe that these two models behave very differently when the tagged particle is exposed to external harmonic potentials or pulling forces. Most importantly, we find that the analytic model has a well defined friction kernel and can be used to extract work, consistent with the microscopic system, while the projection model corresponds to an effective equilibrium model, which cannot be used to extract work. We apply the analysis to two popular non-equilibrium systems, time-delay feedback control and the active Ornstein-Uhlenbeck process. Finally, we highlight that our study could have important consequences for dynamic coarse-graining of non-equilibrium systems going far beyond the linear systems investigated in this manuscript.

\end{abstract}

\maketitle

\section{Introduction}

Coarse graining refers to the procedure in which a complex, high-dimensional microscopic system is substituted with a mesoscopic model featuring significantly fewer degrees of freedom \cite{muller2002coarse,izvekov2005multiscale,peter2009multiscale}. These simplified models, known as coarse-grained (CG) models, offer notable computational efficiency compared to the original system. As a result, they facilitate to connect microscopic time and length scales with macroscopic scales. This process of coarse graining holds significant importance within the realms of statistical physics and computer simulations of soft matter systems \cite{voth2008coarse,brini2013systematic}. 

In recent years, the concept of coarse-graining has been generalized, from purely structural coarse-graining to dynamic coarse-graining, with the goal to derive low-dimensional models with consistent dynamics using the generalized Langevin equation (GLE) \cite{Zwanzig2001,klippenstein2021introducing,schilling2021coarse}. In addition to effective (pair) potentials known from structural coarse-graining \cite{brini2013systematic} the GLE features memory kernels and fluctuating forces to model the frictional interactions and thermal fluctuations in the system.  For equilibrium systems, a manifold of different dynamic coarse-graining techniques has been suggested which can be used to systematically derive such GLEs \cite{Shin2010,li2015incorporation,li2016comparative,lei2016data,jung2017iterative,jung2018generalized,Meyer_2019,wang2020data,D1SM00413A,Meyer2021,ma2021transfer,vroylandt2022likelihood,klippenstein2023_water,she2023data,Obliger2023,Janssen2023}, showing the importance and actuality of the topic. The applicability of these methods to non-equilibrium systems is, however, under strong debate. For instance, most of these techniques are based on the validity of the second fluctuation-dissipation theorem (2FDT) \cite{doi:10.1063/1.5006980,zhu2021effective,zhu2023general,jung2021fluctuation,schilling2021coarse}, which postulates a direct connection between the time-dependent memory kernel in the GLE and the autocorrelation function of the fluctuating forces. However, there is strong evidence that the standard 2FDT is violated in general non-equilibrium systems \cite{Maes2014,maes2014second,Zaccone2018_FDTNESS,netz2018fluctuation,PhysRevE.101.032408,PhysRevE.102.052119, doerries2021correlation,Jung_2022}. This research gap is of critical nature, since there are many examples of non-equilibrium systems, which we would like to analyze in more detail using systematic coarse graining, including nucleation and crystallization \cite{Meyer2021_2}, human cells \cite{PhysRevE.101.032408}, bacteria \cite{bacteria1997} or sperm cells in cervical mucus \cite{tung2017fluid}. Since in such non-equilibrium systems, the emergent structural properties critically depend on the dynamics it would be crucial to develop systematic coarse-graining techniques. Additionally, such methods could be applied far beyond soft mater systems, for example for the analysis of stock markets, ecosystems, or climate, which are inherently out of equilibrium. 

The main objective of this work is therefore to better understand how dynamic coarse-graining techniques can be used to analyze non-equilibrium phenomena and to better analyze transport properties of the reconstructed coarse-grained models. Our analysis is based on an analytically solvable non-equilibrium system \cite{Loos2020,doerries2021correlation}, which, depending on the chosen parameters, can resemble systems featuring time-delayed feedback \cite{loos2014delay,Loos2020,loos2021stochastic,Holubec_2022} or active Ornstein-Uhlenback particles \cite{PhysRevLett.117.038103}. Using this system we have recently highlighted that although the Mori-Zwanzig (MZ) formalism \cite{Zwanzig1961,Mori1965,Zwanzig2001,zhu2021MZstoch,zhu2021effective,zhu2023general} suggests the existence of a 2FDT via the projection operator formalism, the exactly derivable GLE clearly proves that the 2FDT is violated \cite{Jung_2022}. Here, we will significantly extend this work and analyze in detail the two coarse-grained models derived in Ref.~\cite{Jung_2022}. In particular, we will study how they behave in external potentials and under external driving. 

Our results show that although the MZ coarse-grained model properly describes some dynamical properties of the microscopic system, it is not able to capture the intricate non-equilibrium properties of the underlying microscopic system. For example, it does not have the correct response to external forces and it does not flow under the action of an external sawtooth potential. Additionally, we demonstrate that contrary to the exact friction kernel, the MZ memory kernel cannot be easily connected to the frictional forces in the system, thus highlighting the need for the development of more generalized coarse-graining techniques.

Our manuscript is organized as follows. We briefly present the microscopic stochastic model and the dynamic coarse-graining procedures in Section \ref{sec:methods}. Afterwards, we study the behavior of the coarse-grained models under different external conditions: external harmonic potentials (Section \ref{sec:ext_harmonic}), active microrheology (Section \ref{sec:act_microrheology}), response to external forces (Section \ref{sec:linear_response}) and external sawtooth potential (Section \ref{sec:sawtooth}). The results are discussed within the larger context of developing coarse-graining techniques for non-equilibrium systems in Section \ref{sec:discussion}.

\section{Microscopic Model and Dynamic Coarse-Graining}
\label{sec:methods}

In the following, we will recapitulate the main results of Ref.~\cite{Jung_2022} which are necessary basics for the analysis performed in this manuscript.

\subsection{The microscopic model}

The ``microscopic'' system consists of a tagged particle, referred to as ``colloid'', and described by its position $x_0$ and velocity $v_0$. The colloid is coupled to the velocities $ v_i $ of $ N $ ``solvent particles'' via dissipative interactions. The stochastic differential equation (SDE) which defines the system is given by the following set of equations,
\begin{subequations}\label{eq:micro}
\begin{align}
\dot{x}_0(t) &= v_0(t)\\
\dot{v}_0(t) &= F^c(x_0) -\gamma_0 v_0(t) + \sum_{i=1}^N k_i v_i(t) + F^\text{ext}(t),\label{eq:colloid}\\
\dot{v}_i(t) &= - \gamma_i v_i(t) + b_i v_0 + \sqrt{2 k_B T \gamma_i } W_i(t), \quad i> 0, \label{eq:solvent}
\end{align}
\end{subequations}
Here, we have introduced Gaussian white noise $ \langle W_i(t) W_j(t) \rangle = \delta_{ij} \delta(t) $, friction constants $ \gamma_i $, coupling constants $ k_i $ and $ b_i $, as well as the conservative external force $F^c(x_0) = -\frac{\text{d}}{\text{d} x_0} U(x_0),$ and other external forces $F^\text{ext}(t)$.  Throughout the manuscript we set $ \gamma_1 = 1.0 $ and $ k_B T = 1.0 $ which defines the units of the system. While the above SDE should be primarily regarded as a simplified toy model, it can also be viewed as an initial level of coarse-graining for a first-principles system, \gj{similar to dissipative particle dynamics (DPD) \cite{hoogerbrugge1992simulating,espanol1995statistical}}. The friction constants, $ \gamma_i $, and the white noise term $ W_i(t) $ then describe the dissipative and stochastic interaction with a background heat bath of temperature $T.$ The solvent particles could then just represent the first solvation shell around the colloid.

We will study three different variations of the above model, represented by different specific choices for the parameters of the system:
\begin{itemize}
	\item \textbf{EQ}: The system EQ will feature reciprocal interactions, $ b_i = -k_i $, with $ k_1=5 $, $ k_2=2 $, $\gamma_0=0$, \gj{$ \gamma_1 = 1.0 $} and $ \gamma_2=10 $ ($N=2$). The system is therefore in thermal equilibrium and the equipartition theorem is fulfilled, $\langle v_i(0) v_j(0) \rangle = k_B T \delta_{ij} $.
	\item \textbf{FEED}: The non-equilibrium system FEED is characterized by specific non-reciprocal interactions such that the memory kernel has a maximum at time $t>0$, and thus resembles a time-delay feedback mechanism \cite{loos2014delay,Loos2020,loos2021stochastic,Holubec_2022}. The parameters of the system are $ k_1=5=-b_1 $, $ k_2=5=b_2 $, $ \gamma_0=2 $, \gj{$ \gamma_1 = 1.0 $}  and $ \gamma_2=10 $  ($N=2$). This model corresponds to the system NEQ2 of Ref.~\cite{Jung_2022} and inherently violates detailed balance.
	\item \textbf{OU}. The system OU has no coupling of the solvent to the colloid ($b_i=0$) and is thus equivalent to the popular active Ornstein-Uhlenbeck particle model \cite{PhysRevLett.117.038103}. Specifically, we choose $N=1$, $ k_1=5 $, $ \gamma_0=5 $, \gj{and $ \gamma_1 = 1.0 $}, corresponding to NEQ3 in Ref.~\cite{Jung_2022}. Also this model inherently violates detailed balance.
\end{itemize}

\subsection{The exact generalized Langevin equation}

Due to the linearity of the dissipative coupling between colloid and solvent particles, it is easy to derive a generalized Langevin equation (GLE) which describes the non-Markovian dynamics of the colloid after integrating out the solvent particles \cite{Zwanzig2001,netz2018fluctuation,Loos2020,doerries2021correlation,Jung_2022},
 \begin{align}\label{eq:GLE_integrate}
 \dot{x}_0(t) &= v_0(t)\\
\dot{v}_0(t) &= F^c(x_0) - \int_{0}^{t} \text{d}s K^\text{I}(t-s) v_0(s) + \eta^\text{I}(t) + F^\text{ext}(t). \nonumber
\end{align}
In the remainder of this manuscript we will refer to this specific equation as I-GLE.
We find the following explicit expressions for the memory kernel,
\begin{align}
\label{eq:memory_integrate}
K^\text{I}(t) &= \gamma_0 \delta(t) -\sum_{i>0} k_i b_i \exp (-\gamma_i t),
\end{align}
the fluctuating force 
\begin{align}
\label{eq:noise_integrate_full}
\eta^\text{I}(t)&= \sqrt{2 k_B T \gamma_i } \sum_{i>0} \int_{0}^t \text{d}t^\prime k_i  \exp (-\gamma_i t^\prime) W_i(t^\prime),
\end{align}
and thus time-correlation function of the fluctuating force,
\begin{align}
\label{eq:noise_integrate}
C^\text{I}_\eta(t)&=\langle \eta^\text{I}(t) \eta^\text{I}(0) \rangle = k_B T \sum_{i>0}k_i^2  \exp (-\gamma_i t).
\end{align}

\begin{figure}
	\hspace*{-0.5cm}\includegraphics{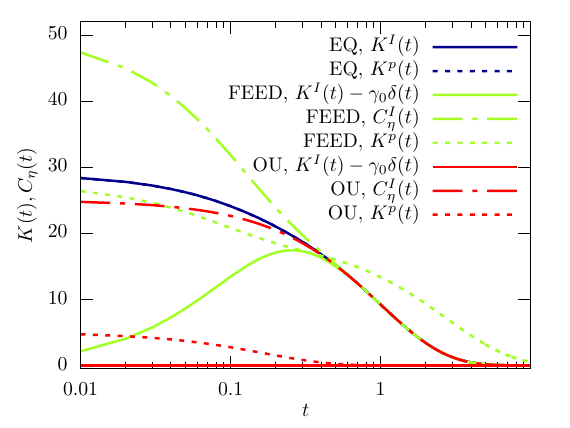}
	\caption{Memory kernel $K(t)$ and noise autocorrelation function $C_\eta(t)$ using the projection method ($p$) or the integration method ($I$) for the three different systems. For system ``EQ'' all curves are overlapping, and $K^x(t)=C^x_\eta(t).$ The projection method fulfills $K^p(t)=C^p_\eta(t)$ for all systems.  }
	\label{fig:memory}
\end{figure}

From Eqs.~(\ref{eq:memory_integrate}) and (\ref{eq:noise_integrate}) it can be concluded that the 2FDT, $ k_B T K^\text{I}(t) = C^\text{I}_\eta(t)$, is only fulfilled for the equilibrium system EQ, but violated in the other two non-equilibrium systems (see Fig.~\ref{fig:memory}). Furthermore, the two specific properties of the non-equilibrium systems become apparent: $K^I(t)$ of FEED has a maximum between $t=0.1 $ and $t=1.0$, and it fulfills $K^I(0)-\gamma_0 \delta(t)=0.$ The memory kernel of the OU system consists, in contrast, only of the instantaneous contribution $K^I(t) = \gamma_0 \delta(t),$ and \gj{the system is only non-Markovian due to the} time-correlation of the fluctuating forces, $C^\text{I}_\eta(t)$ \cite{Jung_2022}.

\subsection{GLE derived via the projection operator technique}

Another systematic way to derive coarse-grained equations of motion (EOM) is via projection operator techniques such as the well-known Mori-Zwanzig formalism \cite{Zwanzig1961,Mori1965,grabert2006projection,Zwanzig2001}. Since the microscopic system in our case is based on stochastic equations, we have applied in Ref.~\cite{Jung_2022} a stochastic extension to the Mori-Zwanzig formalism \cite{zhu2021MZstoch,zhu2023general}. Using these techniques we can derive the following EOM,
\begin{align}\label{eq:GLE_final}
 \dot{x}_0(t) &= v_0(t),\nonumber \\
\dot{{v}}_0(t) &= - \int_{0}^{t} K^{\text{p}}(t-s) {v}_0(s) \text{d}s + \eta^{\text{p}}(t),
\end{align}
with the 2FDT,
\begin{equation}
C_\eta^p(t) = {\langle {\eta}^\text{P}(t){\eta}^\text{P}(0) \rangle} = { \langle v_0(0)^2\rangle  } K^\text{p}(t).
\end{equation}
We have also derived an analytic expression for the memory kernel $ K^{\text{p}}(t)$ \cite{Jung_2022}, which, however, has to be evaluated numerically due to its complexity. In contrast to the analytical I-GLE we therefore find that the projection operator technique immediately postulates the existence of a fluctuation-dissipation theorem. Consequently, with the exception of the EQ system, the memory kernels $K^p(t)$ are different from their exact counterparts $K^I(t),$ as shown in Fig.~\ref{fig:memory}.

\begin{figure}
	\hspace*{-0.5cm}\includegraphics{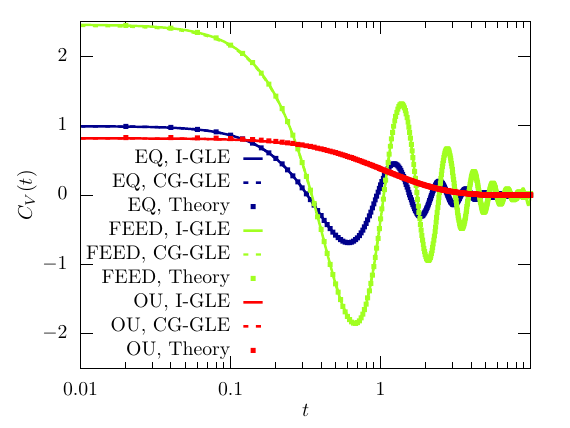}
	\caption{Velocity autocorrelation function $C_V(t)$ as extracted from different coarse-grained models, compared to the theoretical prediction \cite{Jung_2022}. Curves for each system are perfectly overlapping. }
	\label{fig:vacf}
\end{figure}

Several important comments are in order. (i) Per construction, the velocity autocorrelation function (VACF), $C_V(t) = \langle v_0(t) v_0(0) \rangle $ is identical for the dynamics described by the I-GLE and Eq.~(\ref{eq:GLE_final}), as is clearly shown in Fig.~\ref{fig:vacf}. On the level of such pair correlation functions, these EOM are therefore indistinguishable. (ii) The stochastic Mori-Zwanzig formalism \cite{zhu2021MZstoch,zhu2023general} is - strictly speaking - exact. The crucial approximation is when interpreting the quantity $\eta^p(t)$ as a purely stochastic quantity due to the missing knowledge about the initial conditions of the solvent particles. Since there is, however, no possibility to decide which contributions in $\eta^p(t)$ might be stochastic and which might not, this ``uncontrolled'' approximation is the only possible choice to interpret Eq.~(\ref{eq:GLE_final}) as useful coarse-grained model \cite{zhu2023general,Jung_2022}. (iii) While Eq.~(\ref{eq:GLE_final}) has been derived without external forces, it is straightforward to extend the formalism in Ref.~\cite{Jung_2022} to include linear or even non-linear conservative forces, as has been done for a very similar model in Ref.~\cite{jung2023dynamic}, following the lines of Ref.~\cite{vroylandt2022positiondependent}. Formally including general, time-dependent external forces into the formalism is, however, not straightforward and has only very recently been attempted for Newtonian dynamics \cite{koch2023nonequilibrium,netz2023derivation}. \gj{In this manuscript we will therefore make the assumption that the external forces are small, within the linear response regime, and are consequently not significantly changing the memory kernel \cite{jung2021fluctuation,koch2023nonequilibrium}.} These considerations motivate the following generalized Langevin equation,
\begin{align}\label{eq:GLE_final2}
\dot{x}_0(t) &= v_0(t), \\
\dot{{v}}_0(t) &= F^c(t) - \int_{0}^{t} K^{\text{p}}(t-s) {v}_0(s) \text{d}s + \eta^{\text{p}}(t) + F^\text{ext}(t) \nonumber,
\end{align}
which we will refer to as CG-GLE. \gj{It should be noted, that while the result of the projection operator technique, Eq.~(\ref{eq:GLE_final}) is exact, Eq.~(\ref{eq:GLE_final2}) cannot be derived rigorously with the existing projection operator techniques for arbitrary external forces. However, it represents the natural extension of Eq.~(\ref{eq:GLE_final}) for coarse-grained modeling of non-equilibrium systems, which is the goal of this study. Different from Ref.~\cite{shea2023force} in which we have used a similar approach for a complex soft matter system featuring passive probes in active fluids, in the present study we have direct access to the underlying ``ground truth'', i.e. the I-GLE, as reference solution. }

\subsection{Performing GLE simulations}

\gj{In this work aim to study emergent properties of coarse-grained models under the action of external potentials.}
Having \gj{calculated} the memory kernel $K(t)$ and the correlation function of the fluctuating force $C_\eta(t)$ enables us to perform GLE simulations to investigate the behavior of the colloid under the action of different external potentials or non-conservative forces.

To integrate the SDEs (\ref{eq:GLE_integrate}) and (\ref{eq:GLE_final2}) we use the technique proposed in Ref.~\cite{jung2017iterative}. The EOM are therefore discretized using the approach suggested in Ref.~\cite{gronbech2013simple} and the memory convolution integral is discretized using the midpoint rule. The fluctuating force $\eta(t)$ is assumed to be Gaussian (which is, in fact, exact for the present model) and its time-correlation is adjusted using convoluted pseudo random numbers (see Appendix A in Ref.~\cite{jung2017iterative}). 

Based on this technique we perform two kinds of GLE simulations using either the integrated memory kernel $K^I(t)$ (I-GLE) or the projected memory kernel $K^p(t)$ (CG-GLE). The simulations which we will analyze in the following sections only depend on the microscopic system (EQ, FEED or OU) and the applied external forces. For Figs.~\ref{fig:micro} and \ref{fig:response} we additionally explicitly report simulations of the microscopic SDE Eq.~(\ref{eq:micro}) using the discretization derived in Ref.~\cite{Jung_2022}. Although not explicitly shown in every figure, we have validated all results extracted from the I-GLE and found, as expected, perfect agreement with the microscopic simulations.

\section{External Harmonic Potential}
\label{sec:ext_harmonic}

\begin{figure}
	\hspace*{-0.5cm}\includegraphics{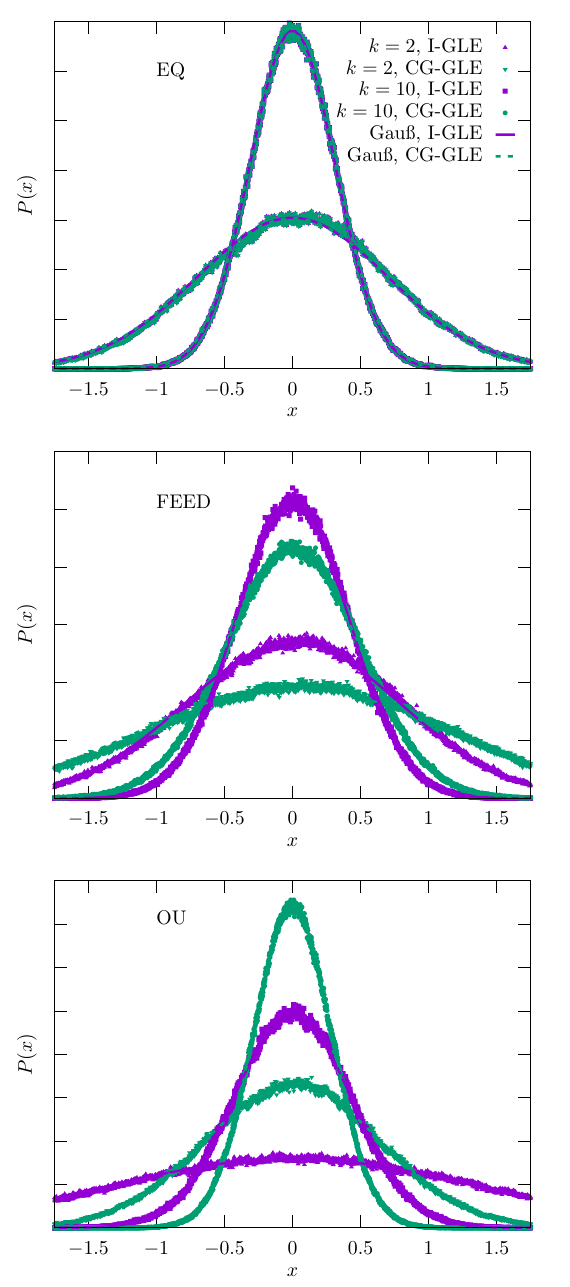}
	\caption{Position probability distribution $P(x)$ as extracted from different coarse-grained models, for harmonic external potentials with strength $k.$ The different subfigures correspond to the three different systems presented in Sec.~\ref{sec:methods}.  }
	\label{fig:harmonic}
\end{figure}

First, we will investigate the behavior of the coarse-grained colloid in a harmonic potential $U(x_0)= \frac{1}{2}k x_0^2$. For equilibrium systems, the probability distribution of the position is trivially given by the Boltzmann distribution, $P(x_0) \propto \exp(-U(x_0) / k_B T),$ and does not depend on the dynamical properties of the system (see Fig.~\ref{fig:harmonic}, EQ).

The situation drastically changes in non-equilibrium systems, in which dynamics can influence structural properties. We clearly observe that the results of I-GLE and CG-GLE differ for the systems FEED and OU (see Fig.~\ref{fig:harmonic}). In fact, the results of the projected GLE indicate that it resembles an equilibrium system, with effective temperature $k_B T^\text{eff} = C_V(0)$, since we find, $P(x_0) \propto \exp(-U(x_0) / k_B T^\text{eff}).$ \gj{To recover the correct $P(x_0)$ in the CG-GLE one would therefore need to use an effective harmonic strength $k^\text{eff}=C_V(0)/\langle x^2 \rangle,$ defined such that  $P(x_0) \propto \exp(- k^\text{eff} x_0^2 / 2 k_B T^\text{eff})$ \cite{shea2023force} in the microscopic simulation.}

\begin{figure}
	\includegraphics[scale=0.9]{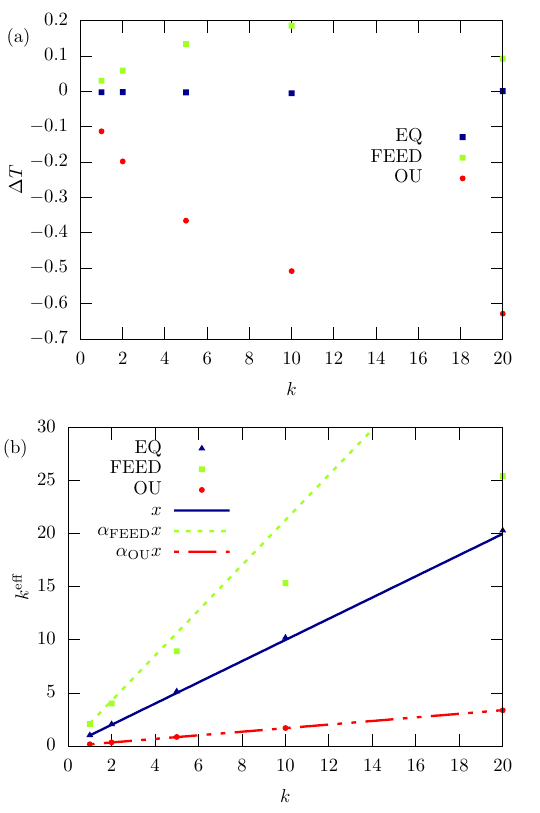}
	\caption{Effective temperature deviation $\Delta T = C_V(0) - C^{k=0}_V(0)$ (a) and effective trap strength ($k^\text{eff}=C_V(0)/\langle x^2 \rangle$) (b) for the three different systems, as extracted from the I-GLE. The dashed-dotted line in (b) is a linear fit to the `OU'' data points, $\alpha_\text{FEED}$ is determined from fitting the linear response in Fig.~\ref{fig:micro}.   }
	\label{fig:harmonic_feed}
\end{figure}

In contrast, the behavior of the exact I-GLE in external potentials is much more complex and shows \gj{emergent} non-equilibrium properties. First, the effective temperature depends on the external potential, as is shown in Fig.~\ref{fig:harmonic_feed}a. For the system FEED, the temperature actually shows a non-monotonic dependence on the harmonic strength $k$, since it first increases and later cools down again. Second, while the distribution $P(x_0)$ remains Gaussian also for the I-GLE, the distribution does not resemble a Boltzmann distribution. To visualize this effect we extract the effective harmonic strength $k^\text{eff}=C_V(0)/\langle x^2 \rangle$. For equilibrium systems we expect and observe $k^\text{eff}=k$ (see Fig.~\ref{fig:harmonic_feed}b). Interestingly, for the OU system we find that $k^\text{eff} = \alpha_\text{OU} k$, indicating that a single renormalization factor $\alpha_\text{OU}$ is sufficient to explain the deviations from equilibrium. This result is in agreement with a similar finding for a passive colloid in suspensions of active particles \cite{shea2023force}. In fact, the active OU particle model has also been proposed as model for passive colloids in active suspensions \cite{wu2000particle}, our results in combination with Ref.~\cite{shea2023force} therefore solidify the intriguing connection between these two systems. \gj{Different from the CG-GLE, in which the effective harmonic strength $k^\text{eff}$ would have to be calculated from $\langle x^2 \rangle$ for each individual $k$, by mapping the system on an effective equilibrium system, this quantity \emph{emerges} in the I-GLE as consequence of the violation of the 2FDT. }

 \gj{Different from the OU system} we find for the FEED system that the linear connection between $k^\text{eff}$ and $k$ is not universal and only applies for small harmonic forces $k<2$. For larger $k$, clear deviations are observable.

The investigation of a colloid in a harmonic trap is an important experimental setup due to the common and simple usage of optical tweezers in similar experiments \cite{Bechinger2022}. The results presented above show that the coarse-graining procedure has important qualitative influence on the behavior of active colloids, or colloids suspended in active baths, trapped in optical tweezers.

\section{Active Microrheology}
\label{sec:act_microrheology}

Another important experimental protocol is pulling the colloid with a constant external force or velocity, thus performing active microrheological experiments \cite{Brau_2007,active_microrheology2009,jung2021fluctuation,jayaram2023effective}. The expectation is that for small and constant external forces $F^\text{ext}$ a linear response regime exists such that the average velocity response $\langle v_0 \rangle = \mu F^\text{ext}.$ The mobility $\mu$ should be connected to the zero-frequency friction coefficient $\zeta$, and thus to the memory kernel, $\mu^{-1} = \zeta = \int dt K(t).$ Active microrheology is therefore one technique to connect the memory kernel to the actual frictional forces acting on the colloid.

\begin{figure}
	\hspace*{-0.5cm}\includegraphics{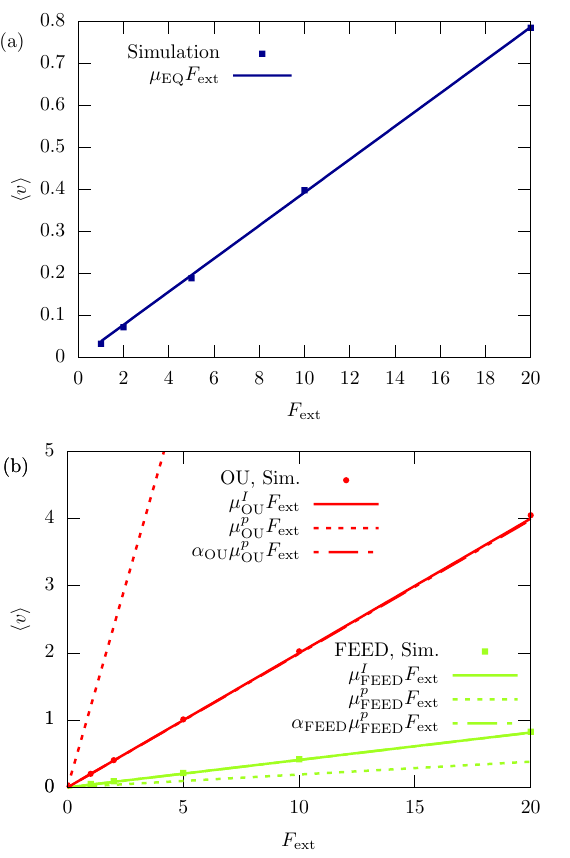}
	\caption{ Average velocity $\langle v \rangle$ as response to a constant external pulling force $F_\text{ext}$, as extracted from microscopic simulations of Eq.~(\ref{eq:micro}), for the equilibrium system (a) and the non-equilibrium systems (b). In addition to the simulation results the figure also features straight lines with gradient equal to the different mobilities extracted from the inverse integrated memory kernels $\mu = \left[\int dt K(t)\right]^{-1}.$ The factor $\alpha_\text{FEED}$ is extracted from fitting the data points ``FEED, Sim.''. }
	\label{fig:micro}
\end{figure}

For the equilibrium system (EQ) we show in Fig.~\ref{fig:micro}a that this relation indeed holds perfectly for the full range of external forces $F^\text{ext}.$ This is not surprising, since the linearity of the microscopic system implies that the linear response regime will span the entire range of possible values for $F^\text{ext}.$

In non-equilibrium systems, we still observe the existence of a linear response regime, enabling us to extract a well defined mobility $\mu_\text{OU}$ (see Fig.~\ref{fig:micro}b). This mobility clearly coincides with the integrated memory kernel, $\mu_\text{OU} = \mu^I_\text{OU} = [\int dt K^I(t)]^{-1},$ thus establishing a clear connection between the memory kernel $K^I(t)$ and the measured friction in the system. This connection, however, does not hold for the CG-GLE, since the integrated memory kernel $[\int dt K^p(t)]^{-1}= \mu^p_\text{OU}  \neq \mu_\text{OU}$. Interestingly, the correction factor $\alpha_\text{OU} = \mu^I_\text{OU} / \mu^p_\text{OU} $ coincides with the rescaling factor derived in the previous section for the effective harmonic strength, consistent with the findings in Ref.~\cite{shea2023force}.  This result suggests that a linear renormalization of forces is sufficient to explain very distinct deviations observed between CG-GLE and the microscopic model. 

For the system FEED, we find similar results, thus strengthen the postulated relationship $\mu_\text{FEED} = \mu^I_\text{FEED}.$ Similar to the above analysis we define a correction factor  $\alpha_\text{FEED} = \mu^I_\text{FEED} / \mu^p_\text{FEED} $ and compare it to the results shown for $k^\text{eff}$ in Fig.~\ref{fig:harmonic_feed}b (see dashed line). While the result does not hold for arbitrarily large values of $k$ due to the non-linear relationship between $k^\text{eff}$ and $k$ we observe that it indeed very well describes the behavior for small $k.$

\section{Linear Response and First Fluctuation-Dissipation Theorem}
\label{sec:linear_response}

In the previous sections we have investigated time-independent steady-state quantities. However, it is also crucial to study the unsteady response to external forces, such  as instantaneous force pulses at $t=0$, $F^\text{ext}(t) = \delta(t).$ From linear response theory we know that the emergent response $\langle v_0(t) \rangle$ is  directly connected to the external force via the response function \cite{forster1975hydrodynamic,jung2021fluctuation}
\begin{equation}
 \langle v_0(t) \rangle = \int_{-\infty}^{\infty} ds \chi(t-s) F^\text{ext}(s) = \chi(t).
\end{equation}
 Importantly, this response function can in equilibrium systems be connected to the velocity autocorrelation function, $\chi(t) = C_V(t) / C_V(0).$ This fundamental relation is called the first fluctuation-dissipation theorem (1FDT) \cite{kubo1966fluctuation}. Since both the response function $\chi(t)$ and the VACF are experimentally measurable quantities, the 1FDT is a fundamental and important relation in statistical physics. 

In non-equilibrium systems, the 1FDT is often violated, and more generalized relationships can be observed \cite{Sciortino2001,Maes2014, Steffenoni2016,maggi2017memory,PhysRevE.101.032408,caprini2021fluctuation,baldovin2022many}. Based on a GLE as presented in Eq.~(\ref{eq:GLE_integrate}), Netz derived a direct relationship between the response function and the memory kernel \cite{netz2018fluctuation},
\begin{equation}\label{eq:response}
\hat{\chi}(\omega) = \frac{1}{ (- \textrm{i} \omega + \hat{K}(\omega)) }.
\end{equation}
Here, we have introduced the one-sided Fourier transformation of a general time-dependent function $F(t)$: $\hat{F}(\omega) = \int_{0}^{\infty} dt F(t) e^{\textrm{i}\omega t}$. Considering that the memory kernel $K^I(t)$ and $K^p(t)$ are different, while the resulting VACFs of the I-GLE and the CG-GLE are identical, this already implies that the 1FDT will be violated in one of our coarse-grained models. In the following, we will calculate the response using the same technique as described in Sec.~III of Ref.~\cite{jung2021fluctuation}.

\begin{figure}
	\hspace*{-0.5cm}\includegraphics[scale=0.9]{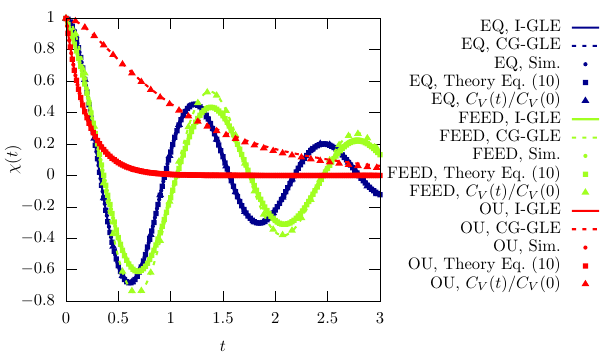}
	\caption{Linear time-dependent response $\chi(t) = \langle v_0(t) \rangle$ to a force impulse, $F_\text{ext}(t)= F_\text{ext} \delta(t),$ at $t=0.$ The response function $\chi(t)$ is compared to the normalized velocity autocorrelation function $C_V(t)/C_V(0),$ and the theoretical prediction, Eq.~(\ref{eq:response}). }
		\label{fig:response}
\end{figure}

The numerical results confirm the theoretical expectation. It can clearly be observed that the equilibrium system fulfills the 1FDT with consistent results for the I-GLE and the CG-GLE (see Fig.~\ref{fig:response}). However, the figure also indicates that in non-equilibrium systems, the I-GLE and the CG-GLE have different time-dependent response to external forces. While the response of CG-GLE is identical to the normalized VACF, and thus the 1FDT is fulfilled, this relationship does not hold for the exact I-GLE. Nevertheless, we find the theoretical prediction Eq.~(\ref{eq:response}) to be confirmed, thus establishing a clear connection between the memory kernel $K^I(t)$ and the response function $\chi(t).$

This result is very significant from a coarse-graining perspective. To have consistent thermodynamic properties in the coarse-grained model, it is of high importance that the memory kernel $K(t)$ can be identified as ``friction kernel''  and thus correctly describes the systematic dynamic interactions between the coarse-grained particles and the surrounding fluid. Our results show that this identification can be done for the I-GLE, but not for the CG-GLE. Furthermore, violation of the 1FDT can be connected to the rate of energy dissipation \cite{harada2005equality}, an essential quantity to characterize non-equilibrium systems and quantify their ability to perform work. 

\section{Sawtooth Potential and Non-Equilibrium Flow}
\label{sec:sawtooth}

The most important property of non-equilibrium systems is their ability to perform work, since this is one of the fundamental building blocks for life. Within the human body, for instance, the molecule adenosine triphosphate (ATP) serves as a perpetual source of energy, ceaselessly harnessed to facilitate muscular contractions, without which the body could not function. This emphasizes the significance of understanding energy dissipation and work of dynamic coarse-grained models, in particular for applications to biological systems.

\begin{figure}
	\hspace*{-0.5cm}\includegraphics{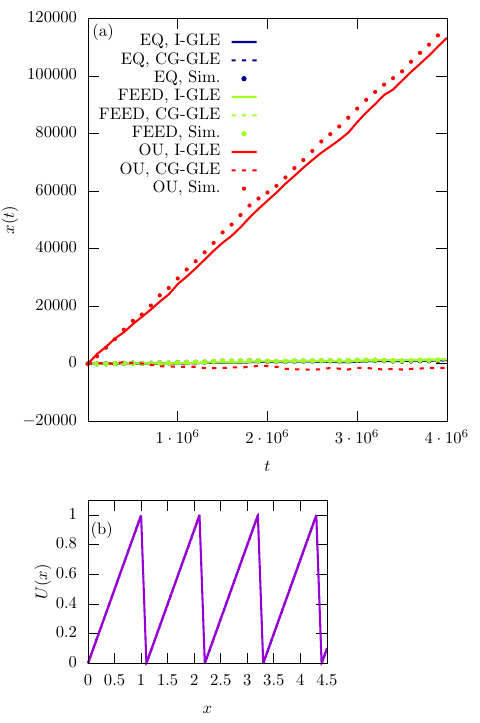}
	\caption{ (a) Time-dependent position $x(t)$ of an individual tracer in a sawtooth potential, extracted from various systems and coarse-grained models. The sawtooth potential is visualized in subfigure (b). }
	\label{fig:sawtooth}
\end{figure}

Here, we investigate a simple protocol to harness energy from non-equilibrium systems: we apply an external, asymmetric sawtooth potential (see Fig.~\ref{fig:sawtooth}b). It is well known that active particles will have a non-zero flow in positive x-direction using such a protocol \cite{bricard2013emergence,PhysRevLett.112.158101,muhsin2022activity}. Using coarse-grained simulations, we observe that, as expected, no flow emerges for the equilibrium system. Interestingly, the same holds for the non-equilibrium FEED system. While this system should, in principle, be able to perform work, the results indicate that other protocols/potentials are required to extract work from such a system (see Fig.~\ref{fig:sawtooth}a). 

The situation is very different for the Ornstein-Uhlenbeck active particles (OU). The colloid definitely has a non-zero flow in the sawtooth potential, consistent with analytical findings in Ref.~\cite{muhsin2022activity}. Fig.~\ref{fig:sawtooth}a also shows that the I-GLE is indeed able to capture this important phenomenon. In contrast, the CG-GLE does not exhibit any flow, consistent with the above finding that \gj{the CG-GLE corresponds to an effective equilibrium model,} the 1FDT is fulfilled and thus the rate of energy dissipation is zero \cite{harada2005equality}.

\section{Conclusions: From toy models to soft matter systems}
\label{sec:discussion}

For coarse-grained modeling it is highly important to validate whether the reduced model accurately describes the properties of the underlying microscopic system. In purely static coarse-graining one is usually interested in structural or thermodynamic properties of equilibrium systems, thus it is not necessarily problematic that dynamic properties are not accurately described. In fact, the speed-up of coarse-grained models is often described as an advantage compared to the atomistic level \cite{brini2013systematic}. In contrast, for many problems in soft matter such as the investigation of transport, the correct description of dynamical properties is crucial \cite{klippenstein2021introducing} even in equilibrium systems although they do not influence emergent structural properties.

The above results go one step further and investigate the importance of modeling the intricate dynamical properties of non-equilibrium systems by correctly identifying frictional forces and active fluctuations. In such situations dynamics can influence all facets of the system, including steady-state probability distributions of particle positions and linear response. We have shown that two coarse-grained models for the same microscopic system, I-GLE and CG-GLE, which yield the same VACF and thus have similar dynamical properties at first sight, are fundamentally different when analyzed with emphasis on several different properties of non-equilibrium systems: violation of fluctuation-dissipation relations, friction or transport. \gj{The \gj{CG-GLE} is based on projection operator techniques and seems to represent an effective equilibrium model which fulfills the standard 2FDT. Contrarily, the I-GLE is able to describe emergent non-equilibrium phenomena such as transport in asymmetric external potentials, which are caused by the violation of the 2FDT. }

Although the results were obtained for a quite simplistic, analytically solvable model we believe that the conclusions are applicable far beyond the scope of the present manuscript to very general non-equilibrium systems. In fact, the complexity of soft matter systems, will, if anything, render precise modeling of the dynamical properties even more crucial. Recently, for example, we have observed very similar phenomenology for a passive particle immersed in an active bath \cite{shea2023force}, which shows similar behavior as the CG-GLE analyzed in the present manuscript. \gj{This can be understood by the orthogonality condition, $\langle v_0(0) \eta^p(t)\rangle=0$, which underlies both the projection operator techniques based on the Mori projector and data-driven coarse-graining tools \cite{shea2023force}.} Other systems already investigated using dynamic coarse-graining include driven actin networks \cite{netz2018fluctuation,abbasi2022non} and cells \cite{PhysRevE.101.032408}. Performing the above analysis for these systems and many other simulations or experiments of soft and biological matter \cite{Gomez-Solano_2015,tung2017fluid} would likely significantly increase our understanding of their fundamental properties and functionality. 

This manuscript therefore highlights the importance of future work in developing reconstructing algorithms of memory kernels in non-equilibrium systems \cite{netz2018fluctuation,vroylandt2022likelihood,abbasi2022non} and the generalization of projection operator techniques for non-equilibrium systems \cite{doi:10.1063/1.5006980,netz2023derivation,koch2023nonequilibrium} to stochastic dynamics. The recently proposed stochastic extension of MZ \cite{zhu2021MZstoch,zhu2021effective,zhu2023general} represents \gj{an important starting point, but clearly needs some extensions to correctly disentangle memory and noise for non-equilibrium dynamics} \cite{Jung_2022}. \gj{It will then be important to investigate whether the same phenomenology described in this paper can be extracted from more realistic microscopic models which are not inherently far away from equilibrium.} 

\section*{Acknowledgements}

The author thanks Friederike Schmid, Bernd Jung and Jeanine Shea for helpful discussions.

\FloatBarrier

\bibliographystyle{unsrtnat}
\bibliography{library_local.bib}

\end{document}